\documentstyle[prl,aps,twocolumn,psfig]{revtex}                

\begin{document}
\draft
\pagestyle{empty}
\wideabs{                                         
\title{Crystal structure of solid Oxygen at high pressure and low temperature}
\author{Federico A. Gorelli$^{1,2}$ \and Mario Santoro$^{1}$ \and Lorenzo Ulivi$^{1,3}$ \and Michael Hanfland$^{4}$}
\address{$^1$ LENS, European Laboratory for Non-linear Spectroscopy and INFM, Largo E. Fermi 2, I-50125 Firenze, Italy\\}
\address{$^2$ Dipartimento di Fisica dell'Universit\`a di Firenze, Largo E. Fermi 2, I-50125 Firenze, Italy\\}
\address{$^3$ Istituto di Elettronica Quantistica, CNR, Via Panciatichi 56/30, I-50127 Firenze, Italy\\}
\address{$^4$ European Synchrotron Radiation Facility, BP220, Grenoble, France\\}
\date{\today}
\maketitle
\begin{abstract}
Results of X--ray diffraction experiments on solid oxygen at low temperature and at pressures up to 10 GPa are presented.
A careful sample preparation and annealing around 240 K allowed very good diffraction patterns in the orthorhombic $\delta$--phase to be obtained.
This phase is stable at low temperature, in contrast to some recent data {[}Y. Akahama {\it et al.}, Phys. Rev. {\bf B64}, 054105 (2001){]}, and transforms with decreasing pressure into a monoclinic phase, which is identified as the low pressure $\alpha$--phase.
The discontinuous change of the lattice parameters, and the observed metastability of the $\alpha$--phase increasing pressure suggest that the transition is of the first order.
\end{abstract}
}             

The magnetic interaction between molecules is of fundamental importance for determining the structural properties of solid oxygen, and, in particular, is responsible for the stabilization of the anti-ferromagnetic $\alpha$ phase.
Indeed, calculations demonstrate that, in absence of magnetic interaction, the lowest energy structure at zero temperature and ambient pressure would be a planar structure, consisting of an hexagonal close packing of parallel oxygen molecules with centers of mass on the planes, and orientation perpendicular to the planes\cite{English74}.
This structure corresponds to the actual oxygen's $\beta$--phase (rhombohedral, space group $R\bar{3}m$), which however is stable only above 23.9 K (up to 43.8 K)\cite{Horl62}.
At lower temperature the stable structure ($\alpha$--phase) is instead monoclinic (space group $C2/m$) \cite{Alikhanov67}, and presents an anti-ferromagnetic order of the molecular spins, which lay in the {\it ab} plane.
The $\alpha$--phase has the same planar arrangement of parallel molecules.
At the $\alpha$--$\beta$ transition, therefore, both structural and magnetic changes occur, and the transition has been described as being of first order, and driven by the magnetic interaction\cite{LeSar88}.
The distortion of the regular hexagons formed by the centers of mass of the molecules in the $\beta$--phase is due to the magnetic interaction and determines an increase of the $b/a$ ratio, which is $1/\sqrt{3} = 0.577$ in the $\beta$--phase and $0.63$ in the $\alpha$--phase at low pressure.

Three different phases have been detected by Raman spectroscopy\cite{Nicol79} in the pressurized solid at room temperature, and their structure has been determined by X--ray diffraction several years ago.
Fluid oxygen crystallizes at room temperature into a rhombohedral structure (stable between 5.5 and 9.6 GPa) which coincides with the low temperature $\beta$--phase\cite{Schiferl81,d'Amour81}, and the phase stable between 9.6 and 9.9 GPa ($\delta$--phase) is orthorhombic, with space group $Fmmm$\cite{Schiferl83}.
The structure of the $\epsilon$--phase, stable above 9.9 GPa at room temperature, is monoclinic \cite{Johnson93,Desgreniers96}, with space group $C2/m$, but the positions of the molecules in the unit cell have not been determined.

The structure of the low temperature monoclinic $\alpha$--phase and of the orthorhombic $\delta$--phase are very closely related and may in principle transform one into the other by a second order transition consisting in a sliding of the basal planes (see fig.~1 of ref.~\onlinecite{Etters85}).
A lattice dynamic calculation predicts instead a first order transition, occurring at about 2.4 GPa\cite{Etters85}.
Experimental information on the oxygen phases at low temperature has been derived only indirectly, and it has been summarized by Desgreniers {\it et al.}\cite{Desgreniers90}.
Indications of the $\alpha$--$\delta$ transition at about 1.0 GPa have been initially reported from low temperature Raman experiments\cite{Meier84b} but other authors describe a phase transition around 0.5 GPa ($\alpha-\alpha'$) and another one around 3 GPa ($\alpha'-\delta'$)\cite{Jodl85}.
Yen and Nicol\cite{Yen87} observe indications of a phase transition at higher pressure and temperature ($\simeq 5$ GPa and 120 K).

Direct X--ray diffraction structural studies at low temperature have not been attempted until recently\cite{Akahama01}.
The authors, increasing pressure along the isotherms at 19, 180 and 240 K, observed that at all the three temperatures the monclinic $\alpha$--phase transforms directly into the $\epsilon$--phase, and therefore confined the range of existence of the $\delta$---phase in the unexplored region above 240 K.
The measured monoclinic distortion is however very small, and possible hysteresis effect may have prevented the determination of the $\alpha$--$\delta$ transition.

We have recently performed infrared spectroscopy studies of oxygen at low temperature and in the pressure range up to 8 GPa.
From the detection of an infrared absorption peak in the frequency region of the fundamental vibration of the oxygen molecule\cite{Gorelli99b}, and, in particular, from its evolution with temperature\cite{Gorelli00}, we demonstrated that solid oxygen is anti-ferromagnetic up to high pressure.
This result has been confirmed recently by a study of the low energy electronic transitions\cite{Santoro01}.

In this paper we present X--ray diffraction results for pressure up to 10 GPa from which the problem of the structure at low temperature can be settled.
Because of a careful annealing procedure, very good diffraction patterns in the orthorombic $\delta$--phase were obtained.
The main conclusion of this study is the observation of the orthorhombic $Fmmm$ phase at temperatures lower than 300 K. 
This phase is stable decreasing $T$, and transforms to the monoclinic $\alpha$--phase decreasing $P$.

We have used a diamond anvil cell (DAC) of the membrane type, loaded condensing high purity liquid oxygen in a sealed vessel.
The cell is fixed to the cold finger of a He flux cryostat.
Temperature is determined by a silicon diode, placed very close to one diamond, with a good reliability ($\pm 1$ K).
Pressure is measured by the shift of the ruby fluorescence wavelength, using the calibration given in ref.~\onlinecite{Ruby}, corrected for measurements at low temperature\cite{RubylowT}.
Sample dimensions ranged from 150 to 170 $\mu$m, with a thickness of about 50 $\mu$m.

Angle dispersive X--ray diffraction has been measured at the ID09 beamline of the European Synchrotron Radiation Facility (ESRF, Grenoble), with a monochromatic beam, ($\lambda = 0.4171 $\AA) and image--plate detection.
The beam reaches the sample through one diamond, forming a square spot with dimension $40 \times 40 \mu$m.
We used a beryllium seat for the second diamond, to obtain a maximum diffraction angle $2\theta \simeq 25$ degrees.
The diffraction patterns were analyzed and integrated by means of the FIT2D computer code, to obtain the one dimensional intensity distribution as a function of the $2\theta$ scattering angle.

Pressurizing fluid oxygen at room temperature into the solid $\beta$--phase usually results in the formation of large crystals in the DAC.
To produce a fine--grained polycrystalline powder, we have caused an abrupt transition of the fluid sample into the $\epsilon$--phase, with a very rapid pressure increase up to about 40--50 GPa.
Subsequently, the $\delta$--$\epsilon$ phase transition, where solid oxygen experiences a large volume change has been crossed a few times in both directions, at a temperature of about 270 K.
With this procedure the preferential orientation of the powder in the $\epsilon$--phase is greatly reduced, as monitored by the almost uniform angular distribution of the intensity along the diffraction rings.

\begin{figure}[bth]
\centerline{\psfig{figure=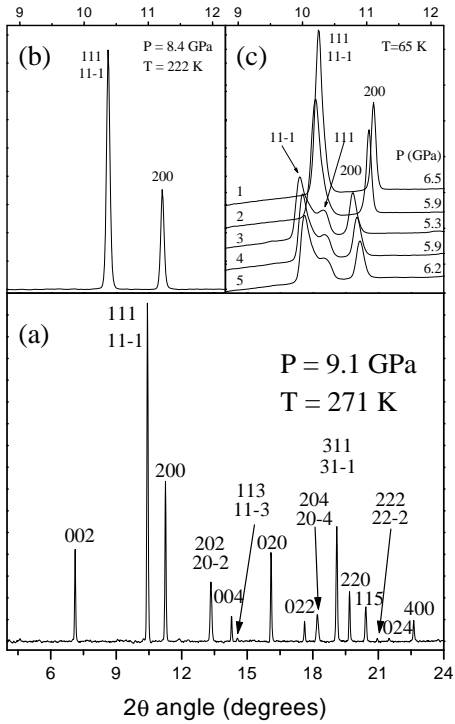,bbllx=0.5cm,bblly=0cm,bburx=21cm,bbury=27cm,width=85mm,angle=0,clip=}}
\caption{(a) Angle dispersive diffraction pattern of oxygen at 9.1 GPa and 271 K ($\lambda$=0.4171\AA) . The indexing correspond to the enlarged monoclinic cell (see text).
(b) The orthorhombic structure persists with cooling of the sample, as demonstrated by the detail of the 111 and 200 reflections at 222 K and 8.4 GPa.
(c) Diffraction patterns obtained during decompression from 6.5 to 5.3 GPa and subsequent compression from 5.3 up to 6.2 GPa at 65 K .
The numbering on the left indicates the time sequence of the measurements, while on the right the pressure is indicated.}
\label{f.dpatt}
\end{figure}

Starting from the $\epsilon$--phase at a temperature of about 270 K, we have slowly decreased the pressure until the $\epsilon$--$\delta$ phase transition was observed.
The diffraction pattern obtained after the transition is represented in fig.~\ref{f.dpatt}(a).
This can be unequivocally indexed on the basis of an orthorhombic cell.
Decreasing slowly the temperature and the pressure, along a stepwise path made up of subsequent isotherms and isobars, we have found that the orthorhombic structure is maintained in a quite large region of the phase diagram.
Indeed, a monoclinic distortion ($\beta \ne 90^{\circ}$) would imply the splitting of the 111, 113, 311, 204, 222 reflections, which instead remain quite narrow (see fig.~\ref{f.dpatt}(b)).
Other cycles performed following different paths in the $P-T$ diagram have demonstrated that a temperature of about 270 K or above is necessary to obtain the annealing of the crystal in the orthorhombic structure in a reasonable time.

Starting from the orthorhombic phase at low temperature, we have collected diffraction patterns during several isothermal decompression runs.
To detect any possible monoclinic deformation we have fitted all the structures leaving the monoclinic angle $\beta$ as a free parameter.
The diffraction patterns obtained during decompression from 6.5 to 5.3 GPa and on compression from 5.3 up to 6.2 GPa at 65 K are shown in fig.~\ref{f.dpatt}(c).
Between 5.9 and 5.3 GPa the diffraction pattern changes drastically with a big shift to lower angle of the 200 line and with a splitting of the 111 line into 111 and $11\bar{1}$.
This is an unambiguous evidence of the transition to the monoclinic phase.
The diffraction pattern at 5.3 GPa can be satisfactorily indexed on the basis of a monoclinic structure with space group $C2/m$, the same of the $\alpha$--phase.
Upon compression at low temperature, the monoclinic structure does not transform again to the orthorhombic one, as testified by the two diffraction patterns measured at the same pressure, 5.9 GPa, before and after the phase transition.
To obtain the orthorhombic structure again the temperature had to be increased up to about 270 K, increasing accordingly the pressure in order to avoid the transformation to the $\beta$ phase.
To our knowledge in the previous spectroscopic experiments\cite{Meier84b,Jodl85,Gorelli99b} the pressure was increased at low temperature, producing the metastability of the monoclinic phase.
The same argument can be used to explain the reason for the missed observation of the phase transition in the diffraction experiment by Akahama et al. \cite{Akahama01}.

\begin{figure}[bth]
\centerline{\psfig{figure=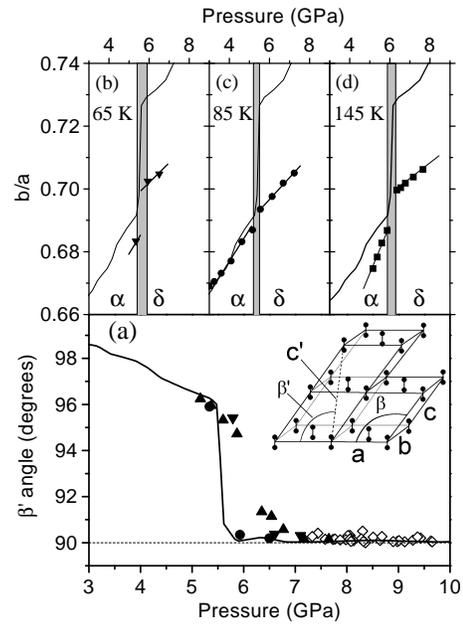,bbllx=0.5cm,bblly=0cm,bburx=20cm,bbury=29cm,width=65mm,angle=0,clip=}}
\caption{(a) Evolution of the angle $\beta'$ along decompression runs.
Empty symbols are at temperatures between 196 and 271, while solid dots, lower triangles and upper triangles are respectively at 65, 102 and 145 K.
(b--d) Evolution with pressure of the ratio $b/a$ of the lattice parameters, at different temperatures, across the $\alpha$--$\delta$ phase transition (solid symbols).
The solid lines in both figures represent the values from lattice dynamic calculation (ref.~\protect\onlinecite{Etters85}), plotted versus rescaled pressure (see text).}
\label{f.dbeta}
\end{figure}

The enlarged monoclinic cell used in the fit of the diffraction patterns (the $abc'$ polyhedron in the inset of fig.~\ref{f.dbeta}(a)) contains 4 molecules and describes as a particular case the orthorhombic one.
In this sense the evolution of the $\beta'$ angle is of particular interest.
This is reported in fig.~\ref{f.dbeta}(a) as a function of pressure along the isotherms that we have performed.
At pressures above 6--7 GPa, depending on temperature, the values of the $\beta'$ angle  are all very close to 90 degrees.
Decreasing pressure, the value of $\beta'$ shows a sudden increase of about 6 degrees, that is the hallmark of the $\delta$--$\alpha$ transition.
The solid line in fig.~\ref{f.dbeta} represents the values of $\beta'$ resulting from lattice dynamics calculation performed at 0 K \cite{Etters85}, plotted after having multiplied the calculated pressure by a factor, to make the calculated transition (2.4 GPa) pressure to coincide with the experimental one (about 5.5 GPa at 85 K).
There is a quantitative agreement with the experiment both for the value of the change of $\beta'$ at the transition and for the pressure slope in the $\alpha$--phase.

As mentioned before, the anti-ferromagnetic order of the $\alpha$--phase produces the distortion of the regular hexagon marking the molecular positions in the $\beta$--phase, and the departure of the ratio $b/a$ from the value $b/a =1/\sqrt{3}$ is a measure of the magnetic interaction between molecules.
It is important therefore to monitor the evolution with pressure of the ratio $b/a$, whose values along the isotherms at 65, 85 and 145 K are reported in fig.~\ref{f.dbeta}(b,c,d), and compared to the lattice dynamics calculation data \cite{Etters85}, plotted, also in this case, after having rescaled the theoretical pressure values.
The ratio $b/a$ increases steadily with pressure, in both the $\alpha$ and $\delta$ phases, confirming the increase of the magnetic interaction between molecules with pressure.

The calculation is quantitatively correct in the $\alpha$--phase, and also the positive sudden increment (if not the size) at the $\alpha$--$\delta$ transition is predicted by the theory.

\begin{figure}[bth]
\centerline{\psfig{figure=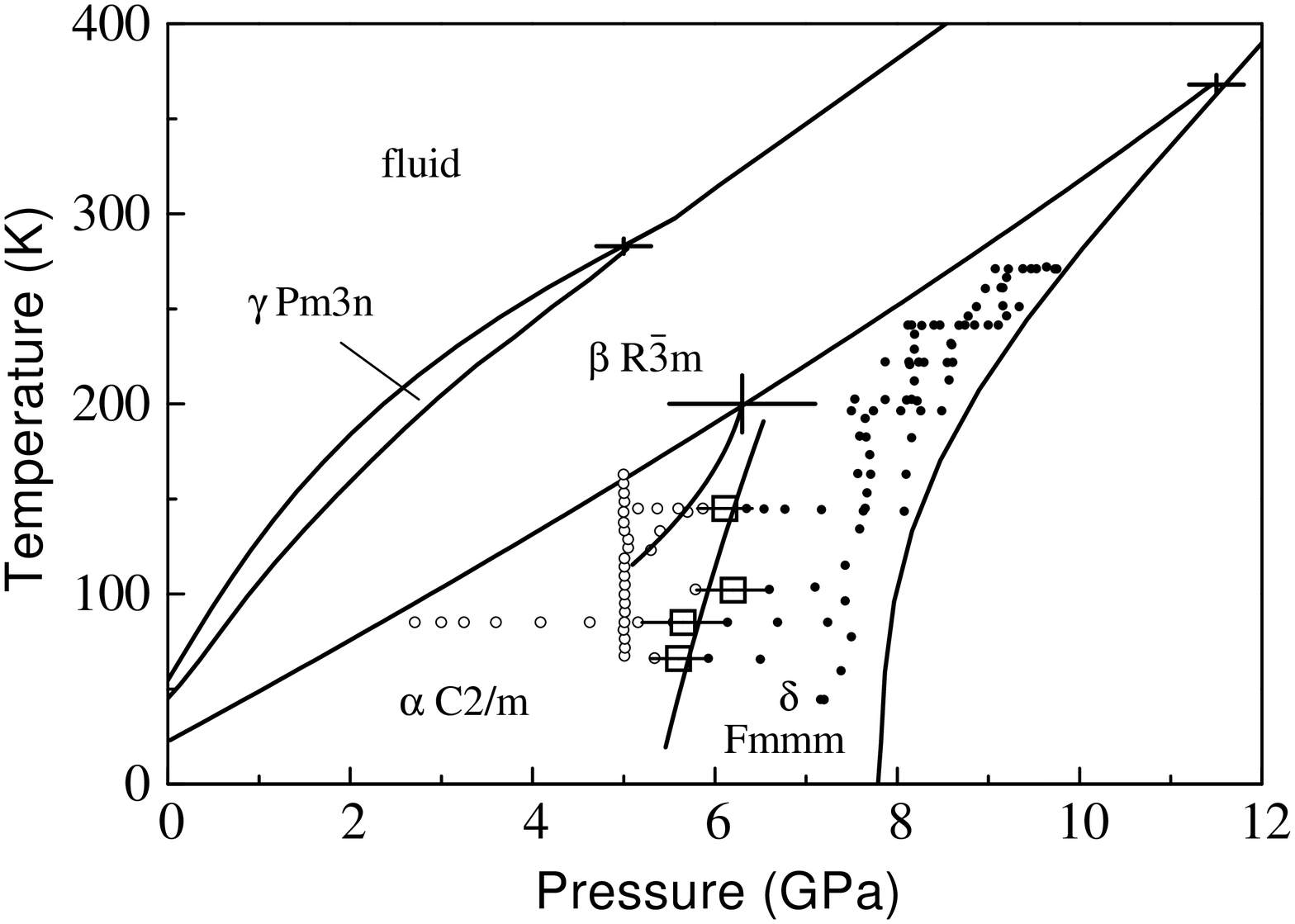,bbllx=0.5cm,bblly=0.0cm,bburx=28cm,bbury=21cm,width=85mm,angle=0,clip=}}
\caption{Phase diagram of solid oxygen. Solids dots and empty circles represent values of $P$ and $T$ where an orthorhombic or monoclinic structure, respectively, has been measured.
The empty squares locate the observed $\delta$--$\alpha$ transition.
The solid line is a possible phase boundary.}
\label{f.phases}
\end{figure}

The main conclusions concerning the low temperature phase diagram are summarized in fig.~\ref{f.phases}.
Full and open circles mark the $P-T$ values where diffraction has been measured, representing respectively the orthorhombic and monoclinic structures.
The line represents a tentative phase boundary between $\alpha$ and $\delta$ phases.
Due to the fact that no evidence of the phase boundary $\delta$'--$\delta$ was observed and that the structure is monoclinic with the same space group of the $\alpha$ phase, $C2/m$, we assign the whole region enclosed by the $\beta$ and $\delta$ phases to the $\alpha$ phase.
The phase boundary identified by Yen and Nicol \cite{Yen87} on the basis of weak extra peaks in the low frequency region of the Raman spectrum is very close to the phase boundary determined in this work.
For this reason we did not remove this phase line, as this might indicate the same $\alpha$--$\delta$ phase transition.

To conclude, our results demonstrate the stability of the orthorhombic $\delta$--phase below room temperature, that can observed when an annealing of the crystal is performed at room temperature or slightly below, but not when a sample is decompressed from the $\epsilon$--phase at low temperature across the $\epsilon$--$\delta$ transition.
In this latter case a mixture of orthorhombic and monoclinic structure is obtained.
The $\alpha$--$\delta$ transition has been observed unambiguously, starting with a good orthorhombic sample and decreasing pressure along four isotherms, appearing with jump in the monoclinic angle $\beta'$ from 90 degrees to about 96 degrees.
The sign and size of this change, and the slope with which the angle $\beta'$ changes with pressure in the $\alpha$--phase (but not the transition pressure) agree in a remarkable way with the theoretical prediction of Etters {et al.} \cite{Etters85}.

The lattice parameters ratio $b/a$ also shows a discontinuous increase passing from $\alpha$--O$_2$ to $\delta$--O$_2$, also predicted by the calculations \cite{Etters85}.

No other phase transitions are detected below 5 GPa.
The phase diagram of solid oxygen therefore has been completely determined, deleting uncertain phase transition boundaries at low temperature.
The sensitivity of the X-ray diffraction technique has been essential to clarify the weakness of previous spectroscopic work, where phase transitions can be detected only by subtle kinks in the pressure slope of mode frequencies.
In addition, the results of this work can furnish a precise recipe to be followed to avoid metastability and obtain a good annealing of an solid oxygen sample.

This research has been supported by the EU contract HPRICT1999-00111.



\end{document}